\documentclass[11pt,aps,preprint,nofootinbib]{revtex4}
\usepackage[active]{srcltx}
\usepackage[utf8]{inputenc}
\usepackage{latexsym}
\usepackage{amsmath}
\usepackage{graphicx}
\usepackage{slashed}
\usepackage{xcolor}

\begin{document}

\title{From Quantum Wires to the Chern-Simons Description of the Fractional Quantum Hall Effect}

\author{Weslei B. Fontana}
\email{weslei@uel.br}
\affiliation{Departamento de F\'isica, Universidade Estadual de Londrina, \\
Caixa Postal 10011, 86057-970, Londrina, PR, Brasil}

\author{Pedro R. S. Gomes}
\email{pedrogomes@uel.br}
\affiliation{Departamento de F\'isica, Universidade Estadual de Londrina, \\
Caixa Postal 10011, 86057-970, Londrina, PR, Brasil}

\author{Carlos A. Hernaski}
\email{hernaski@uel.br}
\affiliation{Departamento de F\'isica, Universidade Estadual de Londrina, \\
Caixa Postal 10011, 86057-970, Londrina, PR, Brasil}

\begin{abstract}

We show the explicit connection between two distinct and complementary approaches to the fractional quantum Hall system (FQHS): the quantum wires formalism and the topological low-energy effective description given in terms of an Abelian Chern-Simons theory. The quantum wires approach provides a description of the FQHS directly in terms of fermions arranged in an array of one-dimensional coupled wires. In this sense it is usually referred to as a microscopic description. On the other hand, the effective theory has no connection with the microscopic modes, involving only the emergent topological degrees of freedom embodied in an Abelian Chern-Simons gauge field, which somehow encodes the collective dynamics of the strongly correlated electrons. The basic strategy pursued in this work is to bosonize the quantum wires system and then consider the continuum limit. By examining the algebra of the bosonic operators in the Hamiltonian, we are able to identify the bosonized microscopic fields with the components of the field strength (electric and magnetic fields) of the emergent gauge field. Thus our study provides a bridge between the microscopic physical degrees of freedom and the emergent topological ones without relying on the bulk-edge correspondence.

\end{abstract}
\maketitle

\textbf{\textit{Introduction}.} 
Quantum Hall (QH) phases of matter have been the focus of intense investigations in the last decades (a nice recent review can be found in \cite{TongQH}). Besides the own interest in their phenomenological properties, the study of QH systems is a source of insightful methods to describe topological phases of matter in general. The nonperturbative nature of the electron-electron interactions turns the microscopic description of topological ordering into a formidable task. The description in terms of effective topological quantum field theories can successfully describe the important low-energy properties of the systems \cite{Wen}. However, since the microscopic degrees of freedom are replaced by effective ones, the connection between the original fermionic Hamiltonian that gives rise to a specific topological phase is obscured in this approach. A new formalism introduced in \cite{Kane} and further developed in \cite{Teo}  reproduces both Abelian and non-Abelian QH phases in terms of a system consisting in a set of one-dimensional wires arranged in parallel. The advantage of this formalism is the direct connection with the microscopic fermionic dynamics. Furthermore, since the two-dimensional problem is turned into a one-dimensional one, usual bosonization techniques can be implemented, allowing to characterize the resulting phases in terms of the edge states and the statistics of the quasiparticle excitations.

Thenceforth, the quantum wires approach has been a subject of great interest and used in several recent investigations. We mention in particular the generalization for other (Abelian and non-Abelian) quantum Hall phases \cite{Cano,Fuji,Furusaki,Stern}, the description of topological insulators \cite{Neupert,Sagi1,Jelena1,Jelena2,Santos,Beri} and superconductors \cite{Sahoo,Park}, constructions of topologically ordered spin liquids \cite{Sela,Meng,Gomes,Patel,Hernaski1}, and also the generalization for higher dimensional topological phases \cite{Meng1,Sagi,Iadecola}. Furthermore, it has been proved a fruitful approach in the study of the recently discovered dualities in three-dimensional field theories - the web of dualities \cite{Karch,Witten}, where the quantum wires can be thought as a discretized version of a 2+1 dimensional fermionic theory. This method made possible explicit derivations of dualities in theories involving fermions \cite{Mross1,Mross2,Hernaski}.

However, in this approach, it remains unclear how to relate the bosonic variables of the one-dimensional quantum wires system with the emergent gauge fields of the 2+1 dimensional topological field theory (except for the bulk-edge correspondence, where there is relation between the emergent gauge field and the boundary degrees of freedom, but not with the bulk degrees of freedom). Such a correspondence would reinforce the consistency of the approaches. By matching the operators in both treatments could also simplify the construction of observables in one framework whose identification is more transparent in the other. Besides, and perhaps more important, the connection of the wires construction with the topological effective theory is the missing link between the starting fermionic model and the low-energy bosonic topological field theory. In this context, if we define a fermionic model in 2+1 dimensions in terms of a set of quantum wires by discretizing one of the spatial dimensions, our discussion indicates a route to 2+1 dimensional bosonization dualities.

In this work we explicitly show how to start from a specific microscopic fermionic Hamiltonian and obtain the low-energy topological field theory for the Laughlin series of the FQHS, given in terms of an emergent Chern-Simons (CS) gauge field. Our strategy is to use a system of quantum wires to describe the Laughlin states, which allows to use 1+1 dimensional bosonization identities. By examining the algebra of the bosonic operators in the Hamiltonian, we are able to identify the bosonized microscopic fields with the components of the field strength (electric and magnetic fields) of the emergent gauge field. The final theory is shown to be the usual CS model with level $m$, where $\nu=1/m$, with $m$ odd, is the filling fraction of the QH phase. Our study provides then a bridge between the microscopic physical degrees of freedom and the emergent topological ones without relying on the bulk-edge correspondence. It is worth to contrast this result with the discussion of \cite{Santos}. There, it is shown that by placing the quantum wires system in an external background electromagnetic field, the low-energy effective action for the background field is a Chern-Simons theory, as it is usually obtained after integrating out the emergent gauge fields. 


\textbf{\textit{Quantum Wires}.} 
The quantum wires approach can be thought as a microscopic description in the sense that the basic degrees of freedom are fermions propagating along an array of one-dimensional system of wires. For each wire, specified by $j=1,...,N$, we associate a pair of chiral complex fermions $\psi_{R/L}^j$, representing the linearized excitations about the Fermi momenta, governed by the free Hamiltonian
\begin{equation}
\mathcal{H}_{0}=\int dx \sum_{j=1}^N -\psi_R^{j\dagger}i\partial_x\psi_R^{j}+\psi_L^{j\dagger}i\partial_x\psi_L^{j}.
\label{0.1}
\end{equation}
This is a sum of one-dimensional uncoupled gapless theories with an electric current that is conserved in each wire, with the corresponding right/left components $J_{R/L}^j=\psi_{R/L}^{j\dagger}\psi_{R/L}^j$. Interactions involving neighboring wires (interwire) that allow electrons to tunnel from one wire to another, then violating charge conservation of the individual wires, have the potential to turn this system into an effectively two-dimensional one.  

We can also introduce many-body forward interactions, which do not open a mass gap in the system. These interactions can be adequately chosen to make some specific operator that plays the fundamental role in the low-energy regime to be perturbatively relevant. For our purposes, it is sufficient to consider the most general current-current intrawire interactions compatible with the symmetries of (\ref{0.1}) (except for Lorentz invariance):
\begin{equation}
\mathcal{H}_{intra}=\int dx\sum_j \frac{\lambda_a}{2} [(J_R^j)^2+(J_L^j)^2]+\lambda_b J_R^j J_L^j.
\label{0.2}
\end{equation}
Upon bosonization, as stated before, the effect of these interactions is just to renormalize the kinetic terms and then are incapable to open a gap in the system. The coupling constants $\lambda_a$ and $\lambda_b$ can be thought then as adjustable parameters of the theory that can be suitably chosen in order to ensure the stability of a topological phase driven by certain interwire interactions.

As discussed in \cite{Teo}, the interwire interactions responsible for generating the Laughlin class of the Abelian FQH is given in the fermionic language by the terrific operator
\begin{equation}
\mathcal{H}_{inter}^{1/m}=-\int dx\sum_j\lambda \left(\psi_{L}^{j+1\dagger}\right)^{\frac{m+1}{2}} \left(\psi^{j+1}_{R}\right)^{\frac{m-1}{2}} \left(\psi_{L}^{j\dagger}\right)^{\frac{m-1}{2}} \left(\psi^{j}_{R}\right)^{\frac{m+1}{2}}+\text{H. c.},
\label{0.3}
\end{equation}
where $m$ is an odd integer. As it has been emphasized in \cite{Neupert}, for the Laughlin phases of FQHS, once we consider interactions that respect conservation of charge, the Haldane criterion, and are restricted to involve only first neighboring wires, then the operator (\ref{0.3}) is essentially unique. It is shown in \cite{Teo} that an appropriate (although explicitly unknown) choice of forward interactions (of density-density type in the bosonic language) can make this operator perturbatively relevant. As we will discuss, the choice (\ref{0.2}) renders this interaction to be irrelevant in the perturbative sense. Then, to be able to get a stable phase at low energies we must consider a strong coupling regime of the $\lambda$ parameter. The advantage of the choice (\ref{0.2}) is the direct identification of the emergent gauge fields in the continuum limit. In the final discussions, we comment on other possible choices for the forward interactions.


\textbf{\textit{Bosonization and the Continuum Limit}.} 
Now we consider the bosonized description of the theory. The bosonization relation is 
\begin{equation}
\psi_{p}^j=\frac{\kappa^j}{\sqrt{2 \pi a}} e^{i( \varphi^j+p\theta^j)},
\label{0.4}
\end{equation}
where $p=R/L=\pm 1$, $a$ is a short-distance cutoff, and $\kappa^j$ are the Klein factors ensuring the anticommutation between fermions of different wires. The boson field $\varphi$ and its dual $\theta$ are related to the usual chiral boson fields $\phi_{L/R}$ through $\varphi^i\equiv \left(\phi^i_R+\phi^i_L+\pi N^i_L\right)/2$ and $\theta^i\equiv \left(\phi^i_R-\phi^i_L+\pi N^i_L\right)/2$, and satisfy the equal-time commutation relations
\begin{equation}
[\theta^i(x),\varphi^j(x')]=i \pi \delta_{ij} \Theta(x-x'),~~~[\theta^i(x),\theta^j(x')]=[\varphi^i(x),\varphi^j(x')]=0,
\label{1.3}
\end{equation}
with $\Theta(x-x')$ being the step function. The Klein factors can be explicitly implemented according to 
\begin{equation}
\kappa^j = e^{i\pi \sum_{i<j}N_L^i +N_R^i},~~~\text{with}~~~ N_p^j =\frac{p}{2\pi}\int dx \partial_x \phi_p^j,
\label{0.5}
\end{equation}
where we are considering periodic boundary conditions along the wires direction.

The bosonized version of the total Hamiltonian,  $\mathcal{H}=\mathcal{H}_{0}+\mathcal{H}_{intra}+\mathcal{H}_{inter}^{1/m}$, reads 
\begin{eqnarray}
\mathcal{H}&=&\int dx\sum_j \frac{v}{2\pi}\left[K(\partial_x\varphi^j)^2+\frac{1}{K}(\partial_x\theta^j)^2 \right]\nonumber\\&-&\int dx\sum_j\frac{\lambda}{\pi}\cos\left[\varphi^{j+1}-\varphi^j-m(\theta^{j+1}+\theta^j)+\pi N^j\right],
\label{1.1}
\end{eqnarray}
where $N^j=N_R^j+N_L^j$ and we have absorbed Klein factors, numerical constants, and factors of $a$ in the coupling constant $\lambda$ \cite{Suple}. In addition, we made a shift in the $\theta$-field around its vacuum configuration, such that all the fields in (\ref{1.1}) have zero vacuum expectation value.
The parameters $v$ and $K$ incorporate the information of the current-current intrawire interactions, and are given by
\begin{eqnarray}
v=\sqrt{\left(1+\frac{\lambda_a}{2\pi}\right)^2-\left(\frac{\lambda_b}{2\pi}\right)^2}~~~\text{and}~~~
K=\sqrt{\frac{1+\frac{\lambda_a}{2\pi}-\frac{\lambda_b}{2\pi}}{1+\frac{\lambda_a}{2\pi}+\frac{\lambda_b}{2\pi}}}.
\label{1.2}
\end{eqnarray}
Upon a rescale of the fields
\begin{equation}
\varphi^j\rightarrow \frac{1}{\sqrt{K}}\varphi^j~~~\text{and}~~~\theta^j\rightarrow \sqrt{K}\theta^j,
\label{1.4}
\end{equation}
which does not affect the commutation rules (\ref{1.3}), we leave the kinetic term in the canonical form: 
\begin{eqnarray}
\mathcal{H}&=&\int dx\sum_j \frac{v}{2\pi}\left[(\partial_x\varphi^j)^2+(\partial_x\theta^j)^2 \right]\nonumber\\&-&\int dx\sum_j\frac{\lambda}{\pi}\cos\left[\frac{1}{\sqrt{K}}(\varphi^{j+1}-\varphi^j)-m\sqrt{K}(\theta^{j+1}+\theta^j)+\pi\sqrt{K}N^j\right].
\label{1.5}
\end{eqnarray}
The quantum dimension of the cosine operator is $\frac{1}{2}\left(\frac{1}{K}+m^2 K\right)$, which indicates that it is irrelevant in the perturbative regime, since the condition $\frac{1}{2}\left(\frac{1}{K}+m^2 K\right)<2$ cannot be satisfied for odd integer $m>1$ and positive $K$. The realization of a two-dimensional gapped phase must then correspond to a strong coupling regime. To explore this possibility, however, we cannot use perturbation theory, but we can still examine the theory (\ref{1.5}) in the continuum limit, where an insightful theory will emerge. 

We expect to make contact with a topological gauge theory emerging in the continuum limit. However, as the Hamiltonian (\ref{1.5}) involves only physical degrees of freedom living at the wires, the relationship with the gauge fields should be through gauge-invariant relations, i.e., $\varphi^i,\theta^j~\Leftrightarrow~E_x, E_y, B$. An useful guide in this direction is to investigate the commutation relations between the operators appearing in the Hamiltonian. With this purpose, we define 
\begin{equation}
\mathcal{O}_1^j\equiv \sqrt{\frac{1}{\pi}}\partial_x\theta^j,~~~\mathcal{O}_2^j\equiv \sqrt{\frac{1}{\pi}}\partial_x \varphi^j,~~~\mathcal{O}_3^j\equiv\frac{1}{\sqrt{K}}(\varphi^{j+1}-\varphi^j)-m\sqrt{K}(\theta^{j+1}+\theta^j)+\pi\sqrt{K}N^j.
\label{1.7}
\end{equation}
By using (\ref{1.3}), we obtain the following equal-time commutation relations:
\begin{equation}
[\mathcal{O}_1^j(x),\mathcal{O}_2^{j'}(x')]=-i  \delta_{j j'}\partial_x\delta(x-x'),
\label{1.8}
\end{equation}
\begin{equation}
[\mathcal{O}_1^j(x),\mathcal{O}_3^{j'}(x')]=i\sqrt{\frac{\pi}{K}} (\delta_{j,j'+1}-\delta_{j j'})\delta(x-x'),
\label{1.9}
\end{equation}
and
\begin{equation}
[\mathcal{O}_3^j(x),\mathcal{O}_2^{j'}(x')]=i m \sqrt{\pi K}  (\delta_{j'+1,j}+\delta_{j j'})\delta(x-x'),
\label{1.10}
\end{equation}
with all other commutators vanishing. 

Proceeding with the continuum limit, we introduce the distance between neighboring wires, $b$, and rescale the fields properly,
\begin{equation}
\varphi^j(t,x) \rightarrow \sqrt{b}\,\varphi(t,x,y)~~~\text{and}~~~\theta^j(t,x) \rightarrow \sqrt{b}\,\theta(t,x,y).
\label{1.11}
\end{equation}  
In addition, we use $\frac{\delta_{j j'}}{b}\rightarrow \delta(y-y')$ and $\delta({\bf r}-{\bf r}')\equiv \delta(x-x')\delta(y-y')$. With this, the above commutation relations become
\begin{equation}
[\mathcal{O}_1({\bf r}),\mathcal{O}_2({\bf r}')]=-i  \partial_x\delta({\bf r}-{\bf r}'),
\label{1.12}
\end{equation}
\begin{equation}
[\mathcal{O}_1({\bf r}),\mathcal{O}_3({\bf r}')]=-i b \sqrt{\frac{\pi}{K}}  \partial_y\delta({\bf r}-{\bf r}'),
\label{1.13}
\end{equation}
and
\begin{equation}
[\mathcal{O}_3({\bf r}),\mathcal{O}_2({\bf r}')]=2 i m \sqrt{\pi K} \delta({\bf r}-{\bf r}')+2 i m \sqrt{\pi K} b \partial_y  \delta({\bf r}-{\bf r}').
\label{1.14}
\end{equation}
These relations show that the operators $\mathcal{O}_{1,2,3}$ produce an algebraic structure very similar to that one between the electric and magnetic fields of the Maxwell-Chern-Simons (MCS) theory \cite{Deser} as we identify
\begin{equation}
\mathcal{O}_1(t,{\bf r})= B(t,{\bf r}),~~~\mathcal{O}_2(t,{\bf r})=E_y(t,{\bf r}),~~~\text{and}~~~\mathcal{O}_3(t,{\bf r})= -b \sqrt{\frac{\pi}{ K}}E_x(t,{\bf r}).
\label{1.14a}
\end{equation}
In this way the above algebra can be written compactly as
\begin{equation}
[B({\bf r}),E_c({\bf r}')]=-i\epsilon_{cd} \partial_d\delta({\bf r}-{\bf r}'), ~~~c,d=x,y,~~~ \epsilon_{xy}=1,
\label{1.15}
\end{equation}
and
\begin{equation}
[E_x({\bf r}),E_y({\bf r}')]=-2 i m \frac{K}{b} \delta({\bf r}-{\bf r}')-2 i m K \partial_y  \delta({\bf r}-{\bf r}').
\label{1.16}
\end{equation}
We see that this sets out an energy scale in the problem: $K/b$. The crucial point is that when the continuum limit is taken, $b\rightarrow 0$, the coupling constants $\lambda_a$ and $\lambda_b$ should be such that $K$ is small enough to ensure $K/b\ll \Lambda$, where $\Lambda\equiv 1/b$. As this energy scale involves only the coupling inside the wires, it corresponds essentially to a gap for excitations propagating in the direction parallel to the wires, which we denote as
\begin{equation}
\frac{K}{b} \equiv \Delta_{\parallel}.
\label{1.17}
\end{equation}
Thus, by taking $K\rightarrow 0$ and $b\rightarrow 0$ keeping fixed $\Delta_{\parallel}$, the last commutator (\ref{1.16}) reduces to 
\begin{equation}
[E_x({\bf r}),E_y({\bf r}')]\approx-2 i m \Delta_{\parallel} \delta({\bf r}-{\bf r}').
\label{1.18}
\end{equation}
In conclusion, the commutators (\ref{1.15}) and (\ref{1.18}) reproduce precisely the algebra of the MCS theory. Furthermore, we note that the coefficient on the right hand side of (\ref{1.18}) involving the odd integer $m$, is related to the Chern-Simons level. This is exactly what is needed for the topological effective field theory to describe the Laughlin series of the FQHS with filling fraction $1/m$. 

We still have to inspect the Hamiltonian (\ref{1.5}) in the continuum limit. In terms of the electric and magnetic fields, according to the identifications in (\ref{1.14a}), it reads
\begin{eqnarray}
\mathcal{H}&=&\int dx \left(\sum_j b\right)\left[ \frac{v}{2}\left( E_y^2+B^2 \right)-\frac{\lambda}{ \pi b}\cos\left( \sqrt{b}\, b \sqrt{\frac{\pi}{K}}E_x\right)\right]\nonumber\\
&=&\int dx dy \left[\frac{v}{2}\left( E_y^2+B^2 \right)+\frac{1}{2}\frac{\lambda b}{ \Delta_{\parallel}}E_x^2+\cdots \right].
\label{1.19}
\end{eqnarray}
A simple dimensional analysis of the quadratic theory shows that $[\lambda]=2$ in mass units, such that $\lambda b$ defines another energy scale in the problem. As it involves only the coupling constant associated with the tunneling of charges between the wires, it corresponds to a gap for excitations propagating in the direction perpendicular to the wires, $\lambda b\equiv\Delta_{\perp}$. This makes evident the necessity for strong coupling ($\lambda\rightarrow\infty$) in order to keep $\Delta_{\perp}$ fixed when $b\rightarrow 0$. Therefore, (\ref{1.19}) reproduces precisely the MCS Hamiltonian when
\begin{equation}
\frac{\Delta_{\perp}}{\Delta_{\parallel}}=v.
\end{equation}
At this point it is instructive to compare the energy scales involved in the problem, which provides valuable insights about the realization of an effectively two-dimensional phase. The continuum limit means that we are considering the regime of energies $E\ll \Lambda$ (large distances). In addition, we have seen that there are other two typical energy scales, $\Delta_{\parallel}$ and $\Delta_{\perp}$, both much below the cutoff $\Lambda$. We then expect that an effectively two-dimensional phase with spatially isotropic motion of charges takes place when the gaps $\Delta_{\parallel}$ and $\Delta_{\perp}$ are comparable, $ \Delta_{\parallel}\sim \Delta_{\perp}$. Otherwise, a strong energetic imbalance like $\Delta_{\parallel}\ll \Delta_{\perp}$ or $\Delta_{\parallel}\gg\Delta_{\perp}$ would produce a highly anisotropic system with an unknown fate.

Thus, we expect that a two-dimensional phase is realized for $v\sim 1$ at the same time as $K\rightarrow 0$. Now it is simple to check that both conditions can be simultaneously met. Let us choose, for example, $v\equiv 1$, which gives a relation between $\lambda_a$ and $\lambda_b$:
\begin{equation}
\left(1+\frac{\lambda_a}{2\pi}\right)=\pm \sqrt{1+\left(\frac{\lambda_b}{2\pi}\right)^2}.
\label{1.20}
\end{equation} 
By considering both constants $\lambda_a$ and $\lambda_b$ positive (repulsive interactions), we get 
\begin{equation}
K(\lambda_b)= \sqrt{\frac{\sqrt{1+\left(\frac{\lambda_b}{2\pi}\right)^2}-\frac{\lambda_b}{2\pi}}{\sqrt{1+\left(\frac{\lambda_b}{2\pi}\right)^2}+\frac{\lambda_b}{2\pi}}},
\label{1.21}
\end{equation}
which shows that $K$ is a decreasing function of $\lambda_b$, with maximum value $K(0)=1$. Thus, by increasing $\lambda_b$ we can make $K$ arbitrarily small. In this way, we see that the conditions on the microscopic parameters for the establishment of the two-dimensional topological phase are mild and can be  easily met. This is in compliance with the low-energy robustness of the topological phase. 

In the case of $v=1$, the Hamiltonian (\ref{1.19}) becomes
\begin{equation}
\mathcal{H}\approx\frac{1}{2}\int d^2r \left(\vec{E}^2+B\right),
\label{1.22}
\end{equation}
with $\vec{E}$ and $B$ satisfying the MCS algebra. The corresponding action reads
\begin{equation}
S=\int d^3x \left( -\frac{1}{4}F_{\mu\nu}F^{\mu\nu}+m \Delta  \epsilon^{\mu\nu\rho}A_{\mu}\partial_{\nu}A_{\rho} \right),
\label{1.23}
\end{equation}
with $\Delta\equiv\Delta_{\parallel}=\Delta_{\perp}$. Therefore, if we consider energies $E\ll \Delta$, the Maxwell term can be ignored leaving only the Chern-Simons term. At the first sight, it seems that we can get rid of both parameters $\Delta$ and $m$ through an appropriate field redefinition. However, this is not the case because, in addition to the effective theory (\ref{1.23}), the identifications in (\ref{1.14a}) imply a certain relation between the total charge of the wires system and the magnetic field (through the magnetic flux). Indeed, the charge of each wire is $\int dx \rho^j(x)$, where the density in the bosonic language reads $\rho^j=\frac{1}{\pi} \partial_x\theta^j(x)$. The  total charge of the system is then given by the integer\footnote{We are assuming that all the wires have the same unit charge and that the system is finite in $y$ direction, having so the geometry of a finite cylinder.} $Q\equiv\sum_j q^j =\sum_j \int dx  \partial_x\theta^j(x)$. With the redefinitions in (\ref{1.4}), this becomes $Q=\sum_j q^j =\sum_j \int dx  \partial_x\theta^j(x) \sqrt{K}$. Next, in the continuum limit, $\theta^j\rightarrow \sqrt{b}\theta$, we obtain
\begin{equation}
Q= \left(\sum_j b\right)  \int dx  \partial_x\theta^j(x) \sqrt{\frac{K}{b}}=\sqrt{\frac{\Delta}{\pi}}\int d^2x B,
\label{1.23a}
\end{equation}
where in the last equality we have used (\ref{1.14a}) and (\ref{1.17}). Therefore, this relation implies flux quantization, i.e., $\int d^2x B \propto \text{integers}$, which means that the gauge fields are compact. This is a reflex of the compact nature of the bosonic wire variables $\theta^j$ and $\varphi^j$. The value of the unit flux itself is arbitrary,  since it can be modified by means of a field redefinition. Incidentally, the energy gap $\Delta$ enters in the Chern-Simons term and in (\ref{1.23a}) in such way that it can be simultaneously eliminated from both expressions through the field redefinition $B\rightarrow \frac{1}{\sqrt{4\pi \Delta}}B$, where we have included numerical factors in the rescaling to bring (\ref{1.23a}) to the usual normalization, that is, $Q=\int d^2x \frac{B}{2\pi}\in \text{\bf Z}$. However, we cannot do the same with the odd integer $m$. It will always be present either in the CS term of the effective action (\ref{1.23}) or in the relation between charge and flux (\ref{1.23a}). This shows that $m$ is truly the only parameter of the microscopic Hamiltonian that manifests itself in the low-energy effective field theory (topological sector). All other parameters, $\lambda_a,\lambda_b$ and $\lambda$ are related to the gap of the system, and hence affect only the dynamics of the collective excitations. This is perfectly compatible with what is  expected in the effective field theory for the FQHS.

By implementing the rescaling as above, $A_{\mu}\rightarrow \frac{1}{\sqrt{4\pi\Delta}} A_{\mu}$, the effective action (\ref{1.23}) becomes
\begin{equation}
S=\int d^3x \left( -\frac{1}{16 \pi \Delta}F_{\mu\nu}F^{\mu\nu}+\frac{m} {4\pi} \epsilon^{\mu\nu\rho}A_{\mu}\partial_{\nu}A_{\rho} \right),~~~\text{with}~\int d^2x \frac{B}{2\pi}\in \text{\bf Z},
\label{1.24}
\end{equation}
making evident how the parameters of the microscopic theory enter in the low-energy effective action. It is interesting to notice that the fact that $m$ is an (odd) integer is consistent with the requirement of the gauge invariance of the partition function under large gauge transformations, inherent to compact fields. 
As a final remark, by defining the system in a finite domain, one can verify the emergence of a chiral massless mode at the boundary described by a chiral Luttinger with fractionalized quantum numbers (charge and statistics). Alternatively, one can find the same edge mode directly in the discrete bosonic model by performing a change of variables from the wires to the links to decouple the boundary wires, according to the discussion in \cite{Teo}.


\textbf{\textit{Discussions}.} 
A crucial point of the work is the identification of the bosonized wire variables $\theta$ and $\varphi$ with the components of the field strength (\ref{1.14a}). Thus it is important to understand whether this is just a fortunate coincidence. To address this question, consider firstly the case of $m=1$, corresponding to an integer quantum Hall system (IQHS), where the interwire interactions (\ref{0.3}) reduce to the form $\psi_L^{j+1 \dagger}\psi_R^j+\text{H. c.}$. In this case, the violation of the electric current conservation in each wire is expressed in terms of the continuity equation $\partial_{0}J_j^0+\partial_x J_j^1=-\delta_jJ_j^2$, where $J_j^{0,1}=J_{R}^j\pm J_L^j$, $\delta_j$ is the discretized derivative operator and the current $J_j^2\sim i \psi_L^{j+1 \dagger}\psi_R^j+\text{H. c.}$. This equation represents the flow of charges into and out of each wire. In the continuum limit, this turns into a 2+1 dimensional conservation law $\partial_{\mu}J^{\mu}=0$, $\mu=0,1,2$, and the current can be written as $J^{\mu}=\bar{\psi}\gamma^{\mu}\psi$, with an appropriate choice of the Dirac matrices. This current, on the other hand, can be parametrized in terms of a gauge field according to $J^{\mu}\sim\epsilon^{\mu\nu\rho}\partial_{\nu}A\rho$. Writing then the components of the current $J^{\mu}$ in the bosonic language, gives us precisely the identifications in (\ref{1.14a}). For general $m$, the same reasoning can be used in terms of the ``dressed" fermions $\widetilde{\psi}_{R/L}^j\equiv(\psi_{L/R}^{j\dagger}\psi_{R/L}^j)^{\frac{m-1}{2}}\psi_{R/L}^j$, in a mechanism somewhat similar to the description of the FHQS in terms of an IQHS but for composite fermions \cite{Jain}. This can be implemented in a precise way via bosonization.

Another point that deserves more comments is about the specific form of the forward interactions in (\ref{0.2}). We emphasize that the most relevant contribution to effective theory (\ref{1.24}), i.e., the Chern-Simons term, carries information of the microscopic theory only by means of the odd integer $m$. The microscopic parameters $\lambda_{a}$ and $\lambda_b$ entering the forward interactions in (\ref{0.2}) are encoded in the energy gap $\Delta$, which includes also $\lambda$. Thus, if we consider more general forward interactions, this will presumably affect only the dynamics of the collective excitations, i.e., the Maxwell term would give place to a more complex structure, but not the topological character captured by the Chern-Simons term. However, a simple identification between the bosonic field theory of the wires and the emergent gauge field as in (\ref{1.14a}) is not guaranteed, involving possibly nonlocal relations. Nevertheless, in the regime of low energies, $E\ll \Delta$, such a microscopic details become unimportant and only the Chern-Simons term plays a crucial role.

Finally, we expect that this type of link between the microscopic degrees of freedom living at the wires and the emergent degrees of freedom encoded in gauge fields, could be extended to the non-Abelian quantum Hall systems as well as to other classes of two-dimensional topological phases of matter. These generalizations are currently under investigation.


\textbf{\textit{Acknowledgments}.} We wish to thank Raul Santos for the enlightening discussions and useful comments on the manuscript. It is also a pleasure to thank Claudio Chamon for the clarifying discussions. We acknowledge the financial support of Brazilian agencies CAPES and CNPq.

\appendix

\section{Supplemental Material for ``From Quantum Wires to the Chern-Simons Description of the Fractional Quantum Hall Effect"}
	
In these notes we provide additional details of the bosonization procedure of the fermionic Hamiltonian responsible for the Laughlin series of the Abelian FQH states considered in the main article
\begin{equation}	
\label{1}
\mathcal{H}=\mathcal{H}_0+\mathcal{H}_{intra} +\mathcal{H}^{1/m}_{inter},
\end{equation}
where,
\begin{equation}
\mathcal{H}_0 = \int\, dx\,\sum_{j=1}^{N}\,-\psi^{j\dagger}_R\,i\partial_x \psi^{j}_R + \psi^{j\dagger}_L\,i\partial_x\psi^{j}_L,
\label{h0}
\end{equation}
\begin{equation}
\mathcal{H}_{intra}=\int \,dx\,\sum_{j=1}^{N} \frac{\lambda_a}{2}\left[(J^{j}_R)^2 + (J^{j}_L)^2\right] + \lambda_b J^{j}_R J^{j}_L,
\label{hintra}
\end{equation}
\begin{equation}
\mathcal{H}_{inter}^{1/m}=-\int dx\sum_j\lambda \left(\psi_{L}^{j+1\dagger}\right)^{\frac{m+1}{2}} \left(\psi^{j+1}_{R}\right)^{\frac{m-1}{2}} \left(\psi_{L}^{j\dagger}\right)^{\frac{m-1}{2}} \left(\psi^{j}_{R}\right)^{\frac{m+1}{2}}+\text{H.c.}.
\label{hinter}
\end{equation}

Following the conventions of \cite{Teo}, we bosonize the Hamiltonian (\ref{1}) according to the fermion-boson mapping:
\begin{eqnarray}
\psi^i_p=\frac{\kappa^i}{\sqrt{2a\pi}}e^{\mathrm{i}\left(\varphi^i+p\theta^i\right)}\label{5},
\end{eqnarray}
with $p=R/L=+1/-1$ and $a$ being a short-distance cutoff. The bosonic fields $\varphi$ and $\theta$ are defined in terms of the chiral boson fields, $\phi_{L/R}$, as $\varphi^i\equiv \left(\phi^i_R+\phi^i_L+\pi N^i_L\right)/2$ and $\theta^i\equiv \left(\phi^i_R-\phi^i_L+\pi N^i_L\right)/2$. The Klein factors $\kappa^i$ are defined in terms of the number operators, $N^i_p=\frac{p}{2\pi}\int{dx\partial_x\phi^i_p}$ as $\kappa^i=\left(-1\right)^{\sum_{j<i}N_L^j+N_R^j}$. They are needed to ensure the anti-commutation relations between fermions associated to different wires. For our purposes the relevant commutation relations are $\left[\theta^i(x),\varphi^j(x^\prime)\right]=\mathrm{i}\pi\delta_{ij}\Theta(x-x^\prime)$, $\left[\theta^i(x),\theta^j(x^\prime)\right]=\left[\varphi^i(x),\varphi^j(x^\prime)\right]=0$, and $\left[N_p^i,\phi_q^j(x)\right]=\mathrm{i}\delta_{ij}\delta_{pq}$, where $\Theta(x-x')$ is the step function. The bosonic field $\theta$ creates fermion-antifermion bound states. Therefore, local polynomials of this field only span the null charge sector of the model. A charged state inside a wire can be created by soliton configuration of this bosonic field. So, the fermion can be seen as coherent bound states, as expressed by (\ref{5}). This physical interpretation can be inferred by noticing that the charge density operator can be shown to be given by $\rho^i(x)=\partial_x\theta^i(x)/\pi$. A unit of charge within the wire then occurs when $\theta$ has a kink where it jumps by $\pi$. Nontrivial topological sectors within the bosonic theory are accounted for by the number operators $N^i_p$, which essentially count the solitons inside a wire.

We proceed to obtain the bosonized version of the three parts $\mathcal{H}_0$, $\mathcal{H}_{intra}$, and $\mathcal{H}_{inter}$ of the Hamiltonian (\ref{1}). Considering the BCH relation $e^Ae^B=e^{A+B+\frac{1}{2}\left[A,B\right]}$, when $\left[A,B\right]$ is a $c$-number, and the proper definitions
\begin{equation}
\psi^{j\dagger}(x)\partial_x\psi^j(x)\equiv\lim_{\epsilon\rightarrow 0}\left[\psi^{j\dagger}(x+\epsilon)\partial_x\psi^j(x)-\left<\psi^{j\dagger}(x+\epsilon)\partial_x\psi^j(x)\right>\right]
\end{equation}
and
\begin{equation}
J^j_p(x)\equiv\lim_{\epsilon\rightarrow 0}\left[\psi^{j\dagger}_p(x+\epsilon)\psi^j_p(x)-\left<\psi^{j\dagger}_p(x+\epsilon)\psi^j_p(x)\right>\right],
\end{equation}
where we are using a point splitting regularization and $\langle\cdots\rangle$ meaning the vacuum expectation value, it follows that
\begin{equation}
\psi^{j\dagger}_p(x)\partial_x\psi^j_p(x)=\frac{ip}{4\pi}\left(\partial_x\left(\varphi^j+\theta^j\right)\right)^2~~~\text{and}~~~
J^j_p=\frac{p}{2\pi}\partial_x\left(\varphi^j+p\theta^j\right).
\end{equation}
Furthermore, using again the bosonization map (\ref{5}) and the BCH relation, we obtain for the operator $\mathcal{O}^{1/m}$ in the integrand of $\mathcal{H}^{1/m}_{inter}$ in (\ref{hinter})
\begin{eqnarray}
\mathcal{O}^{1/m}_{j,j+1}&=&\frac{1}{\left(2a\pi\right)^m}\exp\left[{-i\left[\varphi^{j+1}-\varphi^{j}-m\left(\theta^{j+1}+\theta^{j}\right)+\pi N^{j}\right]-i\pi \left(\frac{m^2-1}{4}\right)+\frac{i\pi}{2}}\right]+\text{H. c.}\nonumber\\
&=&\frac{2}{\left(2a\pi\right)^m}\sin\left[{\varphi^{j+1}-\varphi^j-m\left(\theta^{j+1}+\theta^j\right)+\pi N^j}\right],
\end{eqnarray}
since $m$ is odd, with $N^j=N^j_R+N^j_L$. Using the above results we can write the Hamiltonian (\ref{1}) in the bosonized form
\begin{eqnarray}
\mathcal{H}&=&\int\,dx\,\sum_{j=1}^{N}\frac{1}{2\pi}\left(1+\frac{\lambda_a}{2\pi}-\frac{\lambda_b}{2\pi}\right)\left(\partial_x \varphi^j\right)^2+\frac{1}{2\pi}\left(1+\frac{\lambda_a}{2\pi}+\frac{\lambda_b}{2\pi}\right)\left(\partial_x \theta^j\right)^2\nonumber\\
&-&\frac{\tilde{\lambda}}{\pi}\sin\left[{\varphi^{j+1}-\varphi^j-m\left(\theta^{j+1}+\theta^j\right)+\pi N^j}\right],
\label{sm16}
\end{eqnarray}
where $\tilde{\lambda}$ corresponds to a redefinition of the coupling constant $\lambda$ that includes factors of $a$ and numerical constants that arises in the bosonization of $\mathcal{H}^{1/m}_{inter}$. We can see that the interaction terms of the type $JJ$ only renormalize the kinetic terms. These are the forward scatterings mentioned in the main article. The Hamiltonian (\ref{sm16}) can be written in the form
\begin{eqnarray}
\mathcal{H}&=&\int\,dx\,\sum_{j=1}^{N}\frac{v}{2\pi}\left[K\left(\partial_x \varphi^j\right)^2+\frac{1}{K}\left(\partial_x \theta^j\right)^2\right]\nonumber\\
&-&\frac{\tilde{\lambda}}{\pi}\sin\left[\left(\varphi^{j+1}-\varphi^j\right)-m\left(\theta^{j+1}+\theta^j\right)+\pi N^j\right],\label{10}
\end{eqnarray}
with
\begin{equation}
v=\sqrt{\left(1+\frac{\lambda_a}{2\pi}\right)^2-\left(\frac{\lambda_b}{2\pi}\right)^2}~~~\text{and}~~~ K=\sqrt{\frac{1+\frac{\lambda_a}{2\pi}-\frac{\lambda_b}{2\pi}}{1+\frac{\lambda_a}{2\pi}+\frac{\lambda_b}{2\pi}}}.
\label{sm17}
\end{equation}
We can put the kinetic term of this Hamiltonian into a canonical form by making the rescaling in the fields $\varphi^j\rightarrow \frac{1}{\sqrt{K}}\varphi^j$ and $\theta^j \rightarrow \sqrt{K}\theta^j$, which does not spoil the standard commutation relations. Then, the Hamiltonian assumes the form 
\begin{eqnarray}
\mathcal{H}&=&\int\,dx\,\sum_{j=1}^{N}\frac{v}{2\pi}\left[\left(\partial_x \varphi^j\right)^2+\left(\partial_x \theta^j\right)^2\right]\nonumber\\
&-&\frac{\tilde{\lambda}}{\pi}\sin\left[\frac{1}{\sqrt{K}}\left(\varphi^{j+1}-\varphi^j\right)-m\sqrt{K}\left(\theta^{j+1}+\theta^j\right)+\pi \sqrt{K}N^j\right].
\label{sm18}
\end{eqnarray}

The potential term drives the theory to a non-zero vacuum expectation value of the field $\theta$. We then redefine it as $\theta^j\rightarrow\theta^j+\pi/4 m \sqrt{K}$, to get
\begin{eqnarray}
\mathcal{H}&=&\int dx\sum_{j=1}^{N}\frac{v}{2\pi}\left[\left(\partial_x \varphi^j\right)^2+\left(\partial_x \theta^j\right)^2\right]\nonumber\\
&-&\frac{\lambda}{\pi}\cos\left[\frac{1}{\sqrt{K}}\left(\varphi^{j+1}-\varphi^j\right)-m\sqrt{K}\left(\theta^{j+1}+\theta^j\right)+\pi \sqrt{K}N^j\right],
\end{eqnarray}
where we have renamed the coupling constant $\tilde{\lambda}$ to simply $\lambda$.

The quantum dimension of the cosine operator can be calculated perturbatively by putting this operator in the normal ordered form. Since, $\,e^A\,=:\,e^{A}\,:e^{\frac{1}{2}<AA>}$ and
\begin{equation}
\left\langle \varphi^i \left(x'\right)\varphi^j\left(x\right) \right\rangle=\left\langle \theta^i \left(x'\right)\theta^j\left(x\right) \right\rangle=-\frac{1}{4}\ln\left[{\mu^2 \left(\left(x'-x\right)^2+a^2\right)}\right]\delta_{ij},\label{sm20}
\end{equation}
where $\mu$ is a infrared mass, we obtain
\begin{equation}
\cos\left(A\right)=(\mu a)^{\frac{1}{2}\left(\frac{1}{ K}+m^2 K\right)}:\,\cos\left(A\right):,
\end{equation}
where $A=\frac{1}{\sqrt{K}}\left(\varphi^{j+1}-\varphi^j\right)-m\sqrt{K}\left(\theta^{j+1}+\theta^j\right)+\pi \sqrt{K}N^j$. The quantum dimension of the cosine operator is then $\frac{1}{2}\left(1/K+m^2 K\right)$. We note, in particular, that the dimension reduces to 1 when we set $m=1$ and also turn off the forward interactions, i.e., $K=1$, which corresponds to the bosonization of the mass term of a quadratic fermionic theory.



\end{document}